\documentclass[aps,twocolumn]{revtex4}
\usepackage{amsfonts}
\usepackage{amsmath}
\usepackage{amssymb}
\usepackage{graphicx}

\begin{document}

\title[Short title for running header]{Universal linear relations between
susceptibility and $T_{c}$ in cuprates}
\author{Amit Kanigel, Amit Keren, Arkady Knizhnik, Oren Shafir}
\affiliation{Department of Physics, Technion-Israel Institute of Technology, Haifa 32000,
Israel.}

\begin{abstract}
We developed an experimental method for measuring the intrinsic
susceptibility $\chi $ of powder of cuprate superconductors in the zero
field limit using a DC-magnetometer. The method is tested with lead spheres.
Using this method we determine $\chi $ for a number of cuprate families as a
function of doping. A universal linear (and not proportionality) relation
between $T_{c}$ and $\chi $ is found. We suggest possible explanations for
this phenomenon.
\end{abstract}

\date{\today }
\maketitle

\section{Introduction}

Among the basic properties of a superconductor is its ability to expel a
magnetic field, i.e. the Meissner effect. In all the metallic
superconductors the diamagnetic effect is complete, and below $T_{c}$, the
susceptibility, $\chi $, equals $-1$. In the cuprates high temperature
superconductors (HTSC) the situation is far from being so simple, and there
is growing evidence of samples showing incomplete Meissner effect and even
paramagnetic Meissner effect \cite{PME}. At the same time there is an
accumulation of results showing that the superconducting ground state in
these materials is inhomogeneous \cite{lang}. Therefore, it is possible that
the partial Meissner effect ($\chi <-1$) in the cuprates is an intrinsic
property. This possibility motivated us to perform a comprehensive study of
DC-susceptibility in cuprates. We look for correlations between $T_{c}$ and $%
\chi $, in different HTSC families, and various doping.

It is important to mention that Panagopolous \textit{et al.} measured the AC
susceptibility of La$_{2-y}$Sr$_{y}$ CuO$_{4}$ and HgBa$_{2}$CuO$_{4+\delta }
$ families \cite{PanaPRB99}. However, they were not interested in comparing
the absolute value of $\chi $ between the families and concentrated only on
comparing the temperature dependence of the penetration depth between $\chi $
and $\mu $SR measurements, which resulted in very good agreement.

The difficulty of determining the absolute value of $\chi $ is caused by the
granular nature of the cuprates, and their ability to pin flux very easily.
Consequently, the magnetization in these samples depends very much on the
measurement procedure. For example, cooling a sample in a field, or cooling
in zero field and then applying the field, will result in a different
magnetization. On the other hand, the intrinsic susceptibility of a sample
must be well defined and one should be able to compare different samples.

Therefore, we first develop the condition under which the measurements lead
to the intrinsic susceptibility of the cuprates. The development of these
conditions is based on experience gained while trying to measure the
magnetization of a bundle of lead spheres. Second, we look for correlation
between $T_{c}$ and $\chi $, in different HTSC families with various doping.
Our major finding is a universal linear relation between $T_{c}$ and $\chi $.

Our $\chi $ measurements are done on a set of HTSC families, which are
different in many senses. The different families are La$_{2-y}$Sr$_{y}$CuO$%
_{4}$ (LSCO), YBa$_{2}$Cu$_{3}$O$_{y}$ (YBCO) and its less known
\textquotedblleft cousin\textquotedblright\ (Ca$_{x}$La$_{1-x}$)(Ba$%
_{1.75-x} $La$_{0.25+x}$)Cu$_{3}$O$_{y}$ (CLBLCO) system with 4 different
values of $x$. The CLBLCO \cite{clblco1} system in particular is ideal for
our study due to several interesting properties. Each value of $x$ generates
the full superconductivity dome from the under-doped to the over-doped, and
the maximum $T_{c}$ is $x$ dependent. Thus, each $x$ can be considered as a
superconducting family. For all values of $x$ and $y$ CLBLCO is tetragonal,
so there are no structural transformations that can cause a change in the
volume of the unit cell.

\section{Sample Preparation}

Ceramic samples of LSCO, YBCO and CLBLCO were made by solid-state reaction.
For LSCO, stoichiometric amounts of La$_{2}$O$_{3}$, SrCO$_{3}$ and CuO were
mixed and ground using a ball mill. The mixtures were fired in air for 1-2
days; this was repeated three times. After pelleting, the samples were
sintered in O$_{2}$ for about 64h. The sintering temperature varied between
1100$^{o}$C and 1175$^{o}$C, depending on the Sr level; then the samples
were cooled to room temperature at a rate of 10$^{o}$/h.

For YBCO, the starting materials were Y$_{2}$O$_{3}$, BaCO$_{3}$ and CuO.
The mixture was fired in air at 910$^{o}$C, then pelletized and fired again
at 930$^{o}$C; the last step was then repeated. We also prepared pellets
using YBCO that was supplied by PRAXAIR. This sample is made by combustion
spray pyrolysis; the grains' average size is 3.9 $\mu $m. The pellets of the
two kinds of YBCO were then sintered in O$_{2}$ for 60h at 970$^{o}$C,
cooled at a rate of 10$^{o}$/h down to 510 $^{o}$C, and at a rate of 5$^{o}$%
/h to 410$^{o}$C. The samples were kept at 410$^{o}$C for 5 days and then
cooled down to room temperature at a rate of 10$^{o}$/h.

\begin{table}[tbp]
\label{ybcotable} 
\begin{tabular}{|c|c|c|c|c|}
\hline
$T_c$ & y & $T_r$ & atmosphere & Material \\ \hline\hline
$92(1)K$ & $6.983$ &  &  & PRAXAIR \\ \hline
$86.7(2)K$ & $6.855$ & 530$^oC$ & O$_2$ & Technion \\ \hline
$56.7(2)K$ & $6.549$ & 740$^oC$ & O$_2$ & Technion \\ \hline
$50.5(2)K$ & $6.489$ & 810$^oC$ & O$_2$ & Technion \\ \hline
$40(2)K$ & 6.399 & 840$^oC$ & O$_2$ & PRAXAIR \\ \hline
$20(2)K$ & 6.3 & 580$^oC$ & N$_2$ & PRAXAIR \\ \hline
\end{tabular}%
\caption{Summary of all the YBCO samples and the parameter values used in
their preparation. $T_r$ is the reduction temperature.}
\end{table}

The results presented in this paper for YBCO are from the two types of
samples. No difference can be detected, meaning that the results are not
sensitive to the preparation method of the samples.

The samples oxygen level, y, was then reduced by baking the sample in O$_{2}$
and quenching the samples in liquid nitrogen. For very underdoped samples
the reduction was done in nitrogen atmosphere. The reduction temperatures
are listed in table \ref{ybcotable} . The preparation of the CLBLCO samples
is described elsewhere \cite{clblco1}.

The oxygen level of all the samples was determined by iodometric titration.
In the LSCO samples the deviation of the Oxygen level from 4 is less than
0.005.

The $T_{c}$ of all the samples is determined using resistivity measurements.
In order to compare all the samples, and to overcome the nontrivial problem
of the relation between the chemical doping and $p$, we plot in Fig.~\ref%
{PhaseDiagram} a unified phase diagram using the Presland \textit{et al.}
formula \cite{PreslanPhysicaC91}

\begin{equation}
T_{c}/T_{c,max}=1-82.6(p-0.16)^{2}  \label{TallonP}
\end{equation}%
which relates $T_{c}$ and the holes density $p$.

\begin{figure}[tbp]
\begin{center}
\includegraphics[
width=8.5cm
]{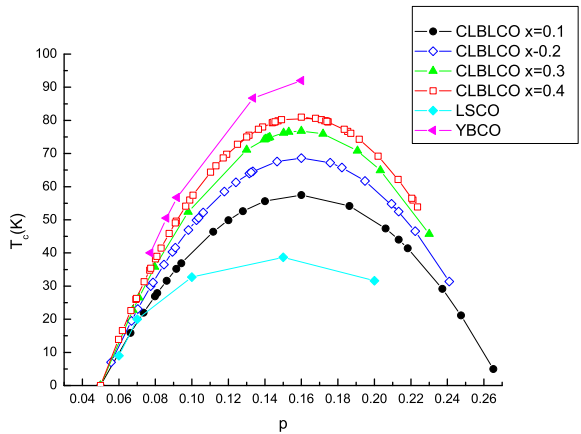}
\end{center}
\caption{(Color online) The phase Diagram of CLBLCO, LSCO, and YBCO after conversion of
chemical doping to hole doping $p$ using Eq. \protect\ref{TallonP}.}
\label{PhaseDiagram}
\end{figure}

\begin{figure}[tbp]
\begin{center}
\includegraphics[
width=8.5cm
]{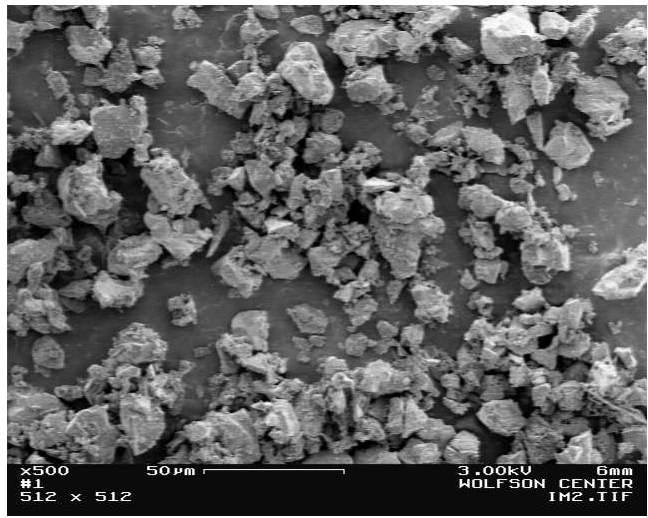}
\end{center}
\caption{SEM picture of a CLBLCO sample with $x=0.4$ and $y=6.983$}
\label{Sem}
\end{figure}

Scanning electron microscopy (SEM) pictures show that the grain sizes in our
YBCO, and CLBLCO for different values of $x$, are of the same order of
magnitude, and that the grains are agglomerates of crystalline whose typical
length is $l\sim 1-10~\mu m$. An example can be seen in Fig. \ref{Sem}.
These properties ensure that the demagnetization factor is similar for the
different families.

\section{Experimental method}

The susceptibility measurements were done using a home built magnetometer
based on a primary coil, two compensating secondary coils, and an extraction
motor. Some results were verified with QD-SQUID at Bar-Ilan University and
with a Cryogenic S600 SQUID magnetometer recently installed in our lab. The
measurements were done in field cool conditions (FC), namely, for field
changes the sample was warmed above $T_{c}$ and cooled down in the new
field. Since we use a superconducting magnet there is always trapped flux in
the magnet leading to a constant shift in the field values. For that reason
the magnetization is measured over a range of positive and negative fields.
The susceptibility is defined by 
\begin{equation}
\chi =\lim_{H\rightarrow 0}\frac{1}{V}\frac{dm}{dH},  \label{chidef}
\end{equation}%
where $m$ is the magnetization obtained from the induced signal at the
secondary coils, $H$ is the external field, and $V$ is obtained from
mass/density. The calibration of the magnetometer is explained below. The
definition of $V$ requires clarification. The problem in powder samples is
to achieve conditions where the volume out of which the field is expelled, $%
V_{sc}$, equals $V$. The zero field cooling condition (ZFC) could result in
a shielding volume $V_{sc}$ which is bigger than $V$ because in certain
geometries Josephson connections can lead to shielding currents enclosing
non-superconducting regions in the sample. The field cool conditions, on the
other hand, lead to the Meissner volume and could result in a $V_{sc}$ which
is smaller than $V$ due to flux pinning. Therefore, our first challenge is
to find the appropriate measurement conditions, where $V$ obtained from
mass/density is exactly the volume out of which the field is expelled for
powders.

\begin{figure}[tbp]
\begin{center}
\includegraphics[
width=8.5cm
]{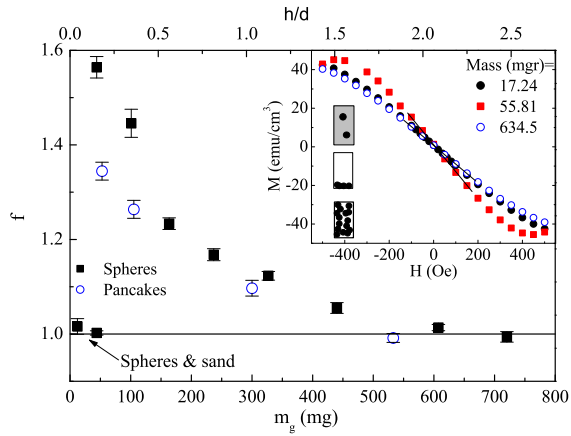}
\end{center}
\caption{(Color online) The volume fraction $f$ of lead, which is the superconducting
volume obtained from $\protect\chi $ measurements, divided by the real
volume taken from mass/density, plotted vs. the mass of samples. In the top
axis we show also the dimensionality of the powder as the height it occupies
in the sample container divided by its diameter. The solid squares represent
sphere shaped grains and open circles represent pancake-shaped grains. In
the inset we show the magnetization curves for three characteristic cases
described in the text.}
\label{Pb}
\end{figure}

In order to gain experience we performed a preliminary experiment with Pb
spheres where the theoretical $\chi $ is well known since Pb is a type I
superconductor, so $\lambda $ is negligible. We used sphere diameter of
0.5~mm, and assumed that $\chi $ is the susceptibility of a single sphere
including the demagnetization factor $(-3/2)$ and obtained $V_{sc}$ $%
=\lim_{H\rightarrow 0}\frac{1}{\chi }\frac{dm}{dH}$. We calibrated the
susceptometer using a few spheres mixed with sand so that they were very
well separated from each other. Raw data are presented in the insert of Fig.~%
\ref{Pb} where we show curves of $M=m/V$ vs $H$ for $3$ samples: (I) the few
Pb spheres mixed with sand ($17.24$mg); (II) a layer of Pb spheres ($55.81$%
mg); and (III) a full container of Pb spheres ($634.5$mg). In all cases a
linear field dependence is observed at low fields. In the first and third
cases we find the same slope at $H\rightarrow 0$, but both are different
from the second case. This means that isolated spheres and a full container
of spheres give the same result.

Our findings in terms of $f=V_{sc}/V$ are summarized in Fig. ~\ref{Pb},
where $f$ is depicted as a function of sample mass, and as a function of
height of spheres in the container ($h$) over its diameter ($d$), on the
lower and upper abscissa, respectively. For a small number the Pb spheres,
which form a 2D layer at the bottom of the container ($h/d<1$), we find $f>1$%
. As the number of spheres increases, the volume they occupy in the
container becomes 3D in nature ($h/d>1$), and $f$ converges to $1$. We
repeated the experiment with \textquotedblleft pancake\textquotedblright\
shaped pieces of lead; the results are qualitatively the same. This leads to
one of the findings of this work. As long as FC conditions prevail and we
use large values of $h/d$, we can safely assume that $V_{sc}=V $. This means
that the magnetic field wanders inside the sample, in between the different
grains, and fills all the empty spaces.

\begin{figure}[tbp]
\begin{center}
\includegraphics[
width=8.5cm
]{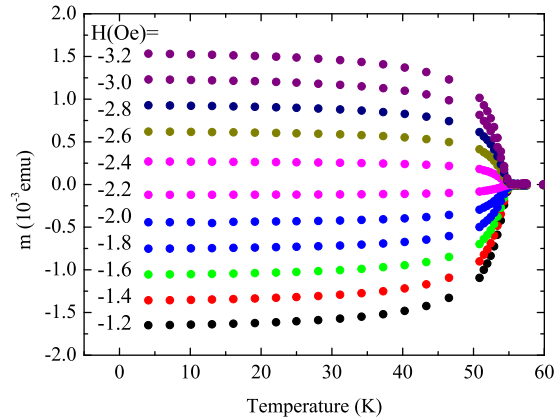}
\end{center}
\caption{(Color online) Field cooled magnetization vs temperature in a variety of fields,
in the small field limit where $H\ll \Phi _{0}/d^{2}$ and $d$ is typical
grain size.}
\label{mvst}
\end{figure}

\begin{figure}[tbp]
\begin{center}
\includegraphics[
width=8.5cm
]{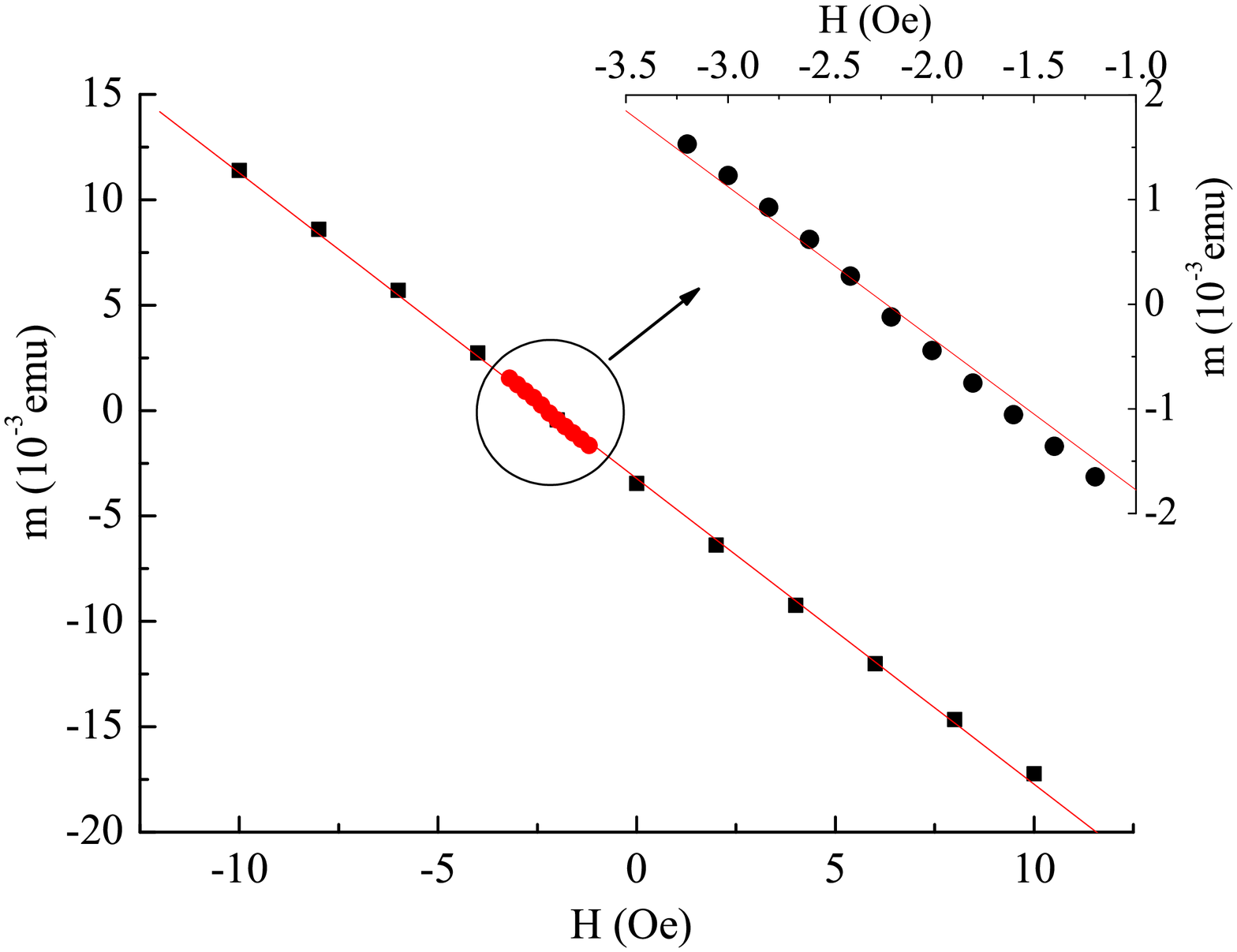}
\end{center}
\caption{(Color online) The zero temperature magnetization after a field cooled is plotted
as a function of applied field. The field scale is shifted due to flux
trapped in the superconducting magnet. Nevertheless, a straight line seems
to fit the data well. Only a zoom in the zero magnetization region, depicted
in the inset, shows that the shift is not identical on the two sides of zero
magnetization. The susceptibility is determined by fitting the data to two
lines, on different sides of zero magnetization, and taking the averaged
slope. }
\label{mvsH}
\end{figure}

In the cuprates $\chi $ of course is unknown, yet it is possible to check if
the experimental conditions developed for Pb apply here as well. For this we
determined the magnetization $m$ in an FC procedure for various applied
fields. We used fields which are small enough that even one flux quanta $%
\Phi _{0}=20~$Oe-$\mu m^{2}$ cannot penetrate our grains (cross section
scale $A\sim 1\mu m^{2}$), namely, $H\ll \Phi _{0}/A=20~$Oe. One such
measurement is shown in Fig. \ref{mvst}. Mostly data in the sub-Oe fields
are completely presented. The magnetization at the lowest temperature as a
function of field is then plotted in Fig. \ref{mvsH}. In this figure a
single line seems to fit the entire field range. However, when zooming in on
the sub-Oe region, which is shown in the inset, a global shift of the line
with respect to the fit is seen between negative and positive fields (due to
bias currents in the power supply). Therefore, we fit $m$ vs. $H$ to two
different lines in a $10$~Oe field range around zero magnetization, and
obtain the susceptibility only from the averaged slope according to Eq.~\ref%
{chidef}. However, outside this $20$ Oe field range a kink in the
magnetization appears which we believe indicates the first vortex that
enters into a grain. Therefore, all data in our experiment were acquired
using this $20$~Oe field range in steps of 1 Oe. We are aware of works
showing a significant non-linear field dependence of the magnetization in
single crystals, especially in very low fields (mOe) \cite%
{Yeshurun2,Yeshurun3}. However, we did not see any deviation from linearity
in our experimental conditions over this 20~Oe range.

\begin{figure}[tbp]
\begin{center}
\includegraphics[
natheight=10.571500in,
natwidth=8.454400in,
height=3.0502in,
width=2.4448in
]{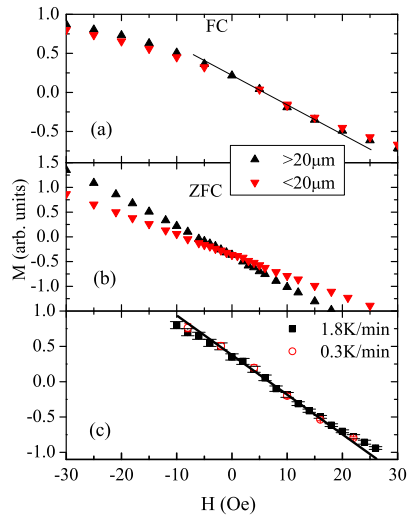}
\end{center}
\caption{(Color online) The magnetization curves for two grain sizes both in (a) FC and (b)
ZFC. (c) depicts measurements for two different cooling rates. }
\label{Cu2OTests}
\end{figure}

To demonstrate that it is the intrinsic susceptibility of the cuprates that
we are measuring, we present four tests. First, we performed the
susceptibility measurements as a function of mass for a CLBLCO sample with $%
T_{c}=42.3$~K in a cylindrical sample holder of $5$~mm inner diameter.
Again, large mass means a 3D cylindric like sample. In contrast the sample
resembles a disk when the mass is small. As can be seen in Fig. \ref%
{ChivsMass}, $\chi $ decreases with increasing mass and saturates. All our
measurements are therefore done with large mass. Second, we use a set of
sieves, and divide the powder grains into two groups: $20\mu m<d<40\mu m$
and $d<20\mu m$ where $d$ is the characteristic size of a grain. We measured 
$\chi $ of these two samples both in FC and ZFC conditions, and the results
are shown in Fig. \ref{Cu2OTests}(a) and (b), respectively. There is hardly
any grain size dependence in the FC measurements, especially when compared
to the ZFC experiment. This indicates that the grain size does not play a
role in determining $\chi $ as long as we use FC procedure.

\begin{figure}[tbp]
\begin{center}
\includegraphics[
width=8.5cm
]{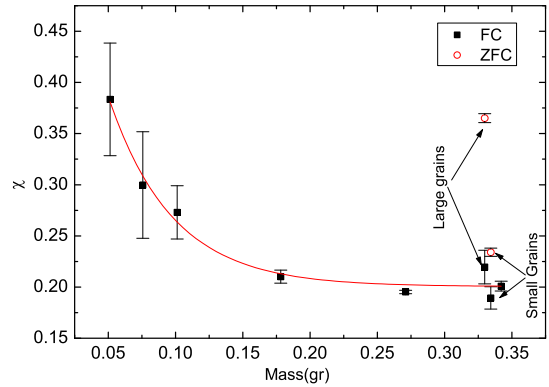}
\end{center}
\caption{(Color online) $\protect\chi $ vs. mass for a CLBLCO sample with $T_{c}=42.3$~K. }
\label{ChivsMass}
\end{figure}

Third, due to the combination of weak flux pinning (compared to low $T_{c}$
superconductors) and high temperatures, the time dependence of the
magnetization can be very complex. Our main interest here is to find the
optimal cooling scheme in a field in order to obtain reproducible
magnetization at base temperatures. We checked the susceptibility of a
sample as a function of the cooling rate. We found, in agreement with
previous works \cite{Yeshurun}, that in FC conditions it is important to
pass through $T_{c} $ slowly. Therefore, in all our measurements we cool the
samples slowly enough so that no difference in the measurements is observed
by cooling them even more slowly. This is demonstrated in Fig. \ref%
{Cu2OTests}(c) where we show that cooling at two different rates does not
vary our result.

\begin{figure}[tbp]
\begin{center}
\includegraphics[
width=8.5cm
]{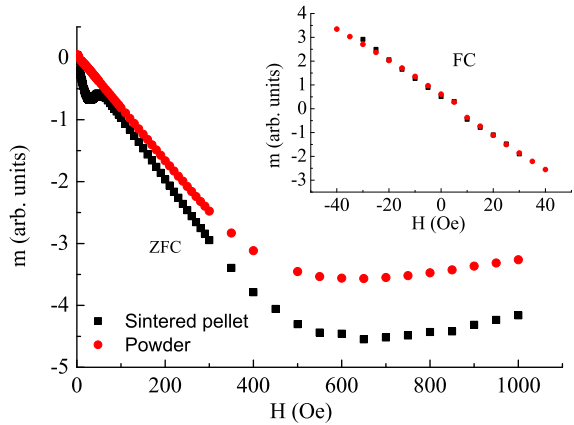}
\end{center}
\caption{(Color online) Magnetization measurements in FC and ZFC conditions of a sintered
pellet and powder obtained by pulverizing the pellet. In FC conditions there
is no difference between the two samples. }
\label{Pellet}
\end{figure}

As a final test we measured the magnetization of a $T_{c}\sim $40K LSCO
sample, first in the form of a sintered pellet and then of the powder after
pulverization. The results are shown in Fig. \ref{Pellet}. In the ZFC
measurements there is a great difference between the magnetization of the
pellet and of the powder. While for the powder we observed the linear
behavior we saw before, for the pellet we find a more complex curve. Up to
20~G the calculated susceptibility is almost -1, indicating a shielding
supercurrent that keeps the entire volume of the sample free of magnetic
flux. Above this field the susceptibility decreases and reaches a value
similar to that of the powder. The difference between the two samples is
only the connection between the grains, those can be described as Josephson
junctions with some average critical field, $H_{Jc1}$. Above this field the
inter-grain links can not support the shielding current and flux penetrates
into the space between the grains and we get local shielding of the grains
as in the powder.

On the other hand, the FC measurements give a different picture. The
magnetization is linear in all fields, both for the pellet and for the
powder. Furthermore, the susceptibility is identical for both samples. This
indicates that the inter-grain links cannot support any Meissner currents at
all. The field is not expelled from the volume in between grains even at
fields below $H_{Jc1}$.

The different behavior of the FC and ZFC measurements in the pellet sample
demonstrate another advantage of our measurement procedure; it is not
sensitive to the connectivity between grains.

We interpret the results of all the above tests as reaching experimental
conditions where small variations of these conditions have no affect on $%
\chi $. Therefore, we believe that our experiments are in the limit where $%
\chi$ is the intrinsic susceptibility of the cuprates.

\section{Results}

\begin{figure}[tbp]
\begin{center}
\includegraphics[
width=8.5cm
]{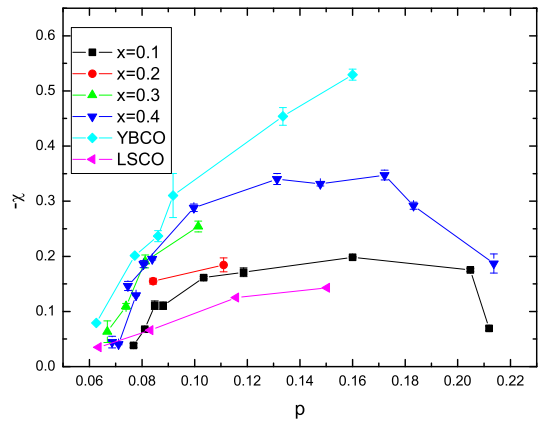}
\end{center}
\caption{(Color online) The negative susceptibility as a function of hole doping $p$ taken
from Eq.\protect\ref{TallonP}.}
\label{VfvsP}
\end{figure}

In Fig.~\ref{VfvsP} we show the FC susceptibilty of all our samples as a
function of $p$, the hole concentration, where $p$ is calculated using Eq.~%
\ref{TallonP}. The curves of $-\chi $ vs. $p$ resemble the phase diagram of
Fig. \ref{PhaseDiagram}, leaving no doubt that $T_{c}$ and $\chi $ are
somehow related.

In Fig. \ref{Uemura}(a) we present $T_{c}$ versus $-\chi $ for all samples.
We find that $T_{c}$ increases linearly (at low doping) with increasing $%
-\chi $, and the linear relation is identical for all families (within
experimental errors). This is the main and theory-independent finding of
this work. It is important to mention that no correlation between $\chi $
and $T_{c}$ was found when the measurements were done in ZFC conditions.

\section{Discussion}

The fact that all the samples obey the same linear relation between $T_{c}$
and $\chi $ is very surprising, given the differences between these cuprate
families. It may be that, because of the complexity of these materials, a
new effective media theory is needed to explain this relation. Nevertheless,
we would like to offer a simpler explanation for our data based on two
experimental observations. On the one hand, it is well known that in the
cuprates there is a universal linear relation between $T_{c}$ and the
inverse in-plane penetration depth squared \cite{UemuraPlot}, known as the
Uemura relation. This relation was revealed by a comprehensive comparison of
the penetration depth, measured by the muon spin relaxation ($\mu $SR)
technique, between different families of HTSC. On the other hand, based on
the growing evidence for inhomogeneity in the cuprates and our observation
that $\chi $ is independent of grain size and connectivity (see Fig.~\ref%
{Pellet} and \ref{Cu2OTests}), it is conceivable that the length scale of
grain sizes observed in Fig.~\ref{Sem} is not the correct grain size.
Therefore, we speculate that the agglomerates seen in Fig.~\ref{Sem} are
made of a very large number of even smaller units stuck together, and that
their number is so large that the size of each one is smaller than the
penetration length, at least in the low doping regime. In this approach the
true effective grain size length scale $a$ would be a parameter to be
determined experimentally.

In type II superconductors such as the HTSC, the penetration length plays an
important role in the susceptibility, and $\chi =-b(1-g(x)/x)$ \cite%
{Shoenberg,Smolyak}, where $x=a/\lambda $. In spherical, plane, and
cylindrical shaped grains $b=3/2,$ $1,$ $1$, and $g(x)$ is the Langevin,
hyperbolic tangent, and the modified Bessel functions, respectively \cite%
{Shoenberg,Smolyak}. Under the assumed long penetration length assumption ($%
a/\lambda <1$) $g(x)/x$ can be expanded to give $\chi =-ca^{2}\lambda ^{-2}$
with $c=1/10$, $1/3$, $1/4$ for the spherical, plane, and cylindrical shaped
cases, respectively. We further speculate that $\chi (\lambda )=O(\lambda
^{-2})$ for all geometries. The anisotropy of the cuprates result in the
replacement $\lambda =1.3\lambda _{ab}$ \cite{FesenkoPhysicaC91}. After
averaging over all grain shapes and sizes we expect 
\begin{equation}
\chi =-\overline{c}\frac{a^{2}}{\lambda _{ab}^{2}}  \label{chitolambda}
\end{equation}%
where $\overline{c}$ is the averaged $c$ and is a number on the order of
unity (including the factor $1.3$).

\begin{figure}[tbp]
\begin{center}
\includegraphics[
width=8.5cm
]{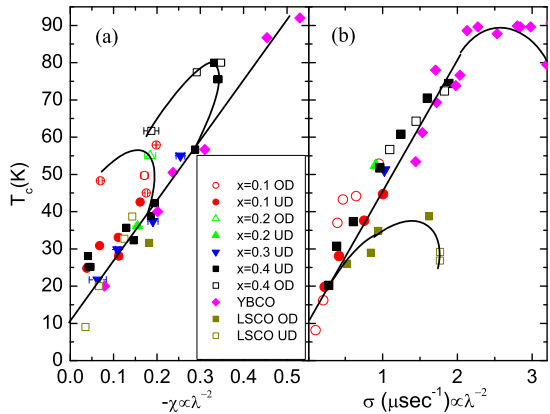}
\end{center}
\caption{(Color online) (a) $T_{c}$ vs $-\protect\chi$ at $T=1.6$~K for various samples of
CLBLCO, YBCO and LSCO. (b) $T_{c}$ vs the muon depolarization rate $\protect%
\sigma$ at $T=1.8$~K for the same CLBLCO samples. Data for YBCO and LSCO are
from Ref. \protect\cite{UemuraPRL89}}
\label{Uemura}
\end{figure}

We also performed $\mu $SR measurements on sintered pellets made from the
same CLBLCO powders, at the Paul Scherrer Institute (PSI) Switzerland, by
field cooling in 3~kOe to 1.8~K. A full account of these measurements in
CLBLCO is given in Ref. \cite{KerenSSC03}, and the YBCO and LSCO data were
taken from Ref.~\cite{UemuraPRL89}. Figure \ref{Uemura}(b) depicts $T_{c}$
versus $\sigma $ for all the samples. Here we used $T_{c}$ from $\mu $SR as
in the original Uemura plot.

A comparison between the two plots reveals interesting information. First,
by comparing the $\mu $SR and susceptibility results we can estimate $a$.
For this we fit both $T_{c}$ versus $\sigma $ and $\chi $ in the underdoped
region to straight lines with offsets $\sigma _{_{0}}$ and $\chi _{_{0}}$.
The solid lines in Fig. \ref{Uemura}(a) and (b) are given by 
\begin{equation}
T_{c}=-K_{\chi }(\chi +\chi _{_{0}}),  \label{Tcsusvschi}
\end{equation}%
and 
\begin{equation}
T_{c}=K_{\mu }(\sigma +\sigma _{_{0}}),  \label{Tcmusrvssigma}
\end{equation}%
respectively, where $K_{\mu }=62(5)$~K-$\mu \sec $ and where $K_{\chi
}=145(5)$~K. We determine $a$ by making $K_{\mu }\sigma $ and $-K_{\chi
}\chi $ agree with each other once they are expressed in terms of $\lambda
_{ab}$. Taking $\sigma =7\times 10^{6}\lambda _{ab}^{-2}$ \cite{MusrBook}
where $\sigma $ is in $\mu \sec ^{-1}$ and $\lambda _{ab}$ in Angstrom, and $%
\chi $ from Eq.~\ref{Tcsusvschi} we obtain 
\begin{equation}
7\times 10^{6}K_{\mu }\lambda _{ab}^{-2}=K_{\chi }\overline{c}a^{2}\lambda
_{ab}{}^{-2}.  \label{Eqfora}
\end{equation}%
Solving this equation with $\overline{c}\lesssim 1$ we find $a\gtrsim 200$%
~nm. This length scale, which is smaller than the typical crystalline size
estimated from SEM, could be due to defects or an intrinsic separation into
domains. The same length scale was also found independently by
ac-susceptibility in YBCO and was ascribed to twinning \cite{PolturakSSC88}.
However, our experiment shows that this is not the origin of $a$ since
CLBLCO and LSCO have no twinning.

Second, there is an offset in both $\chi $ and $\sigma $ so that at $\lambda
_{ab}\rightarrow \infty $ we find $T_{c}\sim 10$ K. This universal deviation
from strict proportionality between $T_{c}$ and $\lambda ^{-2}$ is in
agreement with the measurements of Zuev \emph{et~al.} on YBCO films \cite%
{ZuevCondMat04}. The susceptibility offset could be explained by free spin
that are present in underdoped HTSC and freeze as a spin glass \cite%
{SpinGlass}. The expected susceptibility of paramagnetic spins is given by $%
\chi =4\pi N\mu _{eff}^{2}/(3k_{B}TV)$, where the $4\pi $ is introduced here
since we normalized the susceptibility in Fig.~\ref{Uemura} so that $\chi =-1
$ for a superconductor [instead of $-1/(4\pi )$]. Taking $\mu _{eff}=1.9\mu
_{B}$ per Cu, $T=1.6$~K, $N=3$ spins in a unit cell, and $V$ the volume of a
cell we find $\chi ^{0}\sim 0.1$. However, free spin can not explain the
offset in the $\mu $SR $\sigma $ since they tend to increase $\sigma $
rather than decrease it, namely, with spins $\sigma $ is never zero. A
different explanation for the offset, suggested in Ref.~\cite{ZuevCondMat04}%
, is that $T_{c}\propto \lambda _{ab}^{-2p}$ with $p\sim 1/2$. This power
law is most pronounced in the region $T_{c}<10~$K. This region is out of the
scope of our measurements, but $p\sim 1/2$ at ultra low doping will give an
artificial offset of the $T_{c}$ versus $-\chi $ or $\sigma $ for "normal
doping" ($T_{c}>10$).

Third, panel (b) shows the well known boomerang effect in YBCO and LSCO,
namely, overdoped samples have higher $\sigma $ than underdoped ones with
the same $T_{c}$. In CLBLCO there is an anti-boomerang in both $\mu $SR and $%
\chi $ measurements especially for the $x=0.1$ sample. This is a surprising
result since it means that in overdoped CLBLCO, where the hole concentration
is large, there is in fact a smaller superfluid density ($n_{s}\propto
\lambda _{ab}^{-2}$) than in underdoped samples with smaller hole
concentration.

\section{Conclusions}

We found a universal linear dependence for underdoped HTSC between $T_{c}$
and $\chi $ for different families, with doping as an implicit parameter. A
possible explanation for this dependence is that in underdoped compounds the
penetration depth $\lambda _{ab}$ is longer than an effective grain size
length scale $a$, which is much smaller than the grains size measured using
SEM. In that case $\chi $ is proportional to $a^{2}\lambda _{ab}^{-2}$. By
comparing $T_{c}(\chi )$ and $T_{c}(\sigma )$ we estimate $a\gtrsim 200$~nm.
The amazing aspect of this new grain size is that it is independent of
sample preparation, type of compound, and doping. It appears to be similar
to domain size in ferromagnets which are not determined by the sample size.
In addition, our universal line does not cross the origin in the $T_{c}$, $%
\chi $ plane indicating universal deviation from strict proportionality
between $T_{c}$ and $\lambda ^{-2}$.

\section{Acknowledgment}

We would like to thank the PSI facility for their kind hospitality and
continuing support of this project. We are grateful to Y. Yeshurun for the
use of his QD-SQUID magnetometer and for very helpful discussions. This work
was funded by the Israeli Science Foundation and the Posnansky Research Fund
in High Temperature superconductivity. A. Kanigel would like to thank the
Lady Davis fellowship for financial support.

\end{document}